\documentclass[11pt]{article}
\usepackage{amsmath,amsxtra}
\usepackage{amscd}
\usepackage[dvips]{graphicx}
\usepackage{graphicx}
\usepackage{latexsym}

\usepackage{latexsym}
\usepackage{amsmath}
\begin{document}
\hoffset = -2.4truecm \voffset = -2truecm
\renewcommand{\baselinestretch}{1.2}
\newcommand{\beqa}{\begin{eqnarray}}
\newcommand{\eeqa}{\end{eqnarray}}
\newcommand{\ba}{\begin{eqnarray*}}
\newcommand{\ea}{\end{eqnarray*}}
\newcommand{\mb}{\makebox[10cm]{}\\ }
\newcommand{\pf}{\mbox{pf}}
\date{}

\newtheorem{theorem}{Theorem}
\newtheorem{prop}{Proposition}
\newtheorem{lemma}{Lemma}
\newtheorem{definition}{Definition}

\renewcommand{\theequation}
{\arabic{section}.\arabic{equation}}

\renewcommand{\theequation}
{\arabic{section}.\arabic{equation}}
\title{On a discrete Davey-Stewartson system }
\author{Gegenhasi$^{1,2}$\footnote{gegen@amss.ac.cn}$\quad$ Xing-Biao
Hu$^{1}$\footnote{hxb@lsec.cc.ac.cn} $\quad$ Decio
Levi$^{3}$\footnote{levi@fis.uniroma3.it}
\\
$^{1}$Institute of Computational Mathematics and\\ Scientific
Engineering Computing,
AMSS, Academia Sinica,\\
P.O.\ Box 2719, Beijing 100080, P.R.\ CHINA\\
$^{2}$Graduate School of the Chinese Academy of Sciences, Beijing,
P.R.\ CHINA \\
$^{3}$Dipartimento di Ingegneria Elettronica Universit\'a degli
Studi
Roma Tre \\
Via della Vasca Navale, 84 I00146 Roma Italy } \maketitle
\begin{abstract}
\indent We propose a  differential difference
equation in ${\mathcal R}^1\times {\mathcal Z}^2$ and study it by
 Hirota's bilinear method. This equation has a singular
continuum limit into a system which admits the reduction to the
Davey-Stewartson equation.   The solutions of this discrete DS
system are characterized by  Casorati and Grammian determinants.
Based on the bilinear form of this discrete DS system, we construct
the bilinear B\"{a}cklund transformation which enables us to obtain its
Lax pair.
\end{abstract}
\section{Introduction}
The nonlinear Schr\"odinger equation (NLS) \setcounter{equation}{0}
\begin{equation}
q_t+({\bf v}\bigtriangledown )q-i\sum_{\alpha ,\beta}\frac
{\partial^2\omega}{\partial k_\alpha\partial k_\beta} \frac
{\partial^2q}{\partial x_\alpha\partial x_\beta}-i\omega_1|q|^2q=0
\end{equation}
is the simplest universal model for the slow evolution of the envelope
$q({\bf r}, t)$ of an almost monochromatic wavetrain $\exp (i{\bf
k_0 r}-i\omega (k_0)t)$ in a weakly nonlinear medium of
nonlinear dispersion relation
\begin{equation}
\omega (k)=\omega_0(k)+\omega_1(k)|q|^2+\cdots
\end{equation}
In a $d+1$ dimensional space  this equation has $d+1$ canonical
forms. For $d=1$ they are the "self-focusing NLS"
\begin{equation}
q_t+i(q_{xx}+|q|^2q)=0, \qquad if\quad
\omega_0^{\prime\prime}\omega_1>0\label{NLS1}
\end{equation}
and the  "self-defocusing NLS"
\begin{equation}
q_t+i(q_{xx}-|q|^2q)=0, \qquad if\quad
\omega_0^{\prime\prime}\omega_1<0. \label{NLS2}
\end{equation}
 In the literature, one can find
many results on eqs. (\ref{NLS1}, \ref{NLS2}) and their
discrete versions (see, e.g. \cite{APT,MKNZ} and
references therein).

There are several important generalizations of
the NLS's (\ref{NLS1}, \ref{NLS2}). The best known example
is the Davey-Stewartson (DS) equation \cite{B,DS}, a partial
differential equation in ${\mathcal R}^3$ given by:
\begin{eqnarray}
&&iq_t+\sigma^2(q_{xx}+\sigma^2q_{yy})=-\alpha |q|^2q+q\phi, \label{DS1}\\
&&\phi_{xx}-\sigma^2\phi_{yy}=2\alpha (|q|^2)_{xx}, \label{DS2}
\end{eqnarray}
where $\alpha^2=1, \sigma^2=\pm 1$. In the hyperbolic case
$\sigma^2=1$,  the system (\ref{DS1}, \ref{DS2}) is called DSI while
DSII corresponds to the elliptic case $\sigma^2=-1$. In the
following we will focus on the DSI equation. The DSI equation
(\ref{DS1}, \ref{DS2})  is a reduction of the system
\begin{eqnarray}
&&iq_t+(q_{xx}+q_{yy})=-\alpha q^2r+q\phi, \label{DS3}\\
&&-ir_t+(r_{xx}+r_{yy})=-\alpha qr^2+r\phi, \label{DS4}\\
&&\phi_{xx}-\phi_{yy}=2\alpha (qr)_{xx}, \label{DS5}
\end{eqnarray}
obtained by requiring that  $r=q^*$.
By the  variable transformation $\partial_x=\frac 1{\sqrt
2}(\partial_X+\partial_Y), \partial_y=\frac 1{\sqrt
2}(\partial_X-\partial_Y), \phi =\alpha qr+\alpha\psi$, the system
(\ref{DS3}--\ref{DS5}) may be transformed into
\begin{eqnarray}
&&iq_t+(q_{XX}+q_{YY})=\alpha q\psi, \label{DS6}\\
&&-ir_t+(r_{XX}+r_{YY})=\alpha r\psi, \label{DS7}\\
&&\psi_{XY}=\frac 12(\partial_X^2+\partial_Y^2)(qr). \label{DS8}
\end{eqnarray}
Integrable discrete versions for the DS equations have not been much
studied yet although much work has been done on the discrete NLS
equation \cite{APT}. A few partial results have been presented by
Nijhoff and Konopelchenko in \cite{K,K1,N}.  The purpose of this
paper is to propose a new discrete integrable system of equations
which can be considered as a discrete version for the DSI system
(\ref{DS6}--\ref{DS8}).

Let us consider the following system
\begin{eqnarray}
&&iv_t+\alpha_1 e^{u_{n-1}+u_{n+1}-2u}v_{n-1}+\alpha_2
e^{u_{k-1}+u_{k+1}-2u}v_{k+1}-(\alpha_1+\alpha_2)v=0,\label{N1}\\
&&-iw_t+\alpha_1 e^{u_{n-1}+u_{n+1}-2u}w_{n+1}+\alpha_2
e^{u_{k-1}+u_{k+1}-2u}w_{k-1}-(\alpha_1+\alpha_2)w=0,\label{N2}\\
&&z_1-z_1e^{u_{n+1,k+1}+u-u_{k+1}-u_{n+1}}+z_2v_{k+1}w_{n+1}=0\label{N3},
\end{eqnarray}
where $\alpha_1,\alpha_2,z_1$ and $z_2$ are constants. In eqs.
(\ref{N1}--\ref{N3}) and in the following we always use a
simplified notation for $f(n,k,t)$.  We  write explicitly a
discrete independent variable only when it is shifted from its position. For
example,
$$
f\equiv f(n,k,t), \quad f_{n+1}\equiv f(n+1, k,t),\quad
f_{k+1}\equiv f(n,k+1,t),\quad f_{n+1,k-1}\equiv f(n+1,k-1,t).
$$

Let us now show that eqs. (\ref{N1}--\ref{N3}) may be
thought of as a discrete version of the DSI system
(\ref{DS6}--\ref{DS8}). Let us set
\begin{equation*}
\alpha_1=\frac 2{\epsilon^2},\quad \alpha_2= \frac 2{\delta^2},\quad
z_1=\frac 1{\epsilon\delta},\quad z_2=-\frac 14\alpha ,\quad
\epsilon n= X,\quad \delta k=Y,
\end{equation*}
and expand the dependent fields in power series around $\delta = 0$ and $\epsilon
= 0$,
\begin{eqnarray*}
&&v_{k+1}=v(n,k)\equiv q(n\epsilon , (k+1)\delta)=q(X,Y+\delta
)=q+\delta q_Y+\frac {\delta^2}2q_{YY}+\cdots \\
&&w_{n+1}=w(n+1,k)\equiv r((n+1)\epsilon , k\delta )=r(X+\epsilon
,Y)=r+\epsilon r_X+\frac {\epsilon^2}2r_{XX}+\cdots \\
&&u_{n+1}=u+\epsilon u_X+\frac {\epsilon^2}2u_{XX}+\cdots \\
&&u_{n+1,k+1}=u+\epsilon u_X+\delta u_Y+\frac
{\epsilon^2}2u_{XX}+\epsilon\delta u_{XY}+\frac
{\delta^2}2u_{YY}+\cdots\\
&&\qquad\qquad\qquad \cdots\cdots
\end{eqnarray*}
Then the continuous limit of eqs.
(\ref{N1}--\ref{N3}) gives
\begin{eqnarray}
&&iq_T +(q_{XX}+q_{YY})=-2q(\partial_X^2+\partial_Y^2)u, \label{N4}\\
&&-ir_T +(r_{XX}+r_{YY})=-2r(\partial_X^2+\partial_Y^2)u, \label{N5}\\
&&u_{XY}=-\frac 14\alpha qr, \label{N6}
\end{eqnarray}
where $i\partial_T=i\partial_t-\frac 2\epsilon\partial_X+\frac
2\delta\partial_Y $.  Under the transformation
$\psi =-2\alpha(\partial_X^2+\partial_Y^2)u, T\longrightarrow t$, the system (\ref{N4}--\ref{N6}) reduces to
the DSI system (\ref{DS6}--\ref{DS8}) .

In the following we will study eqs.
(\ref{N1}--\ref{N3}) using Hirota bilinear method. By the dependent
variable transformation
\begin{equation}
u=\ln F,\qquad v=e^{-i(\alpha_1+\alpha_2)t}G/F,\qquad
w=e^{i(\alpha_1+\alpha_2)t}H/F \label{TR}
\end{equation}
eqs. (\ref{N1}-\ref{N3}) are transformed into the bilinear form
\begin{eqnarray}
&&[iD_t+\alpha_1e^{-D_n}+\alpha_2e^{D_k}]G \cdot F=0,\label{bf1}\\
&&[iD_t+\alpha_1e^{-D_n}+\alpha_2e^{D_k}]F \cdot H=0,\label{bf2}\\
&&z_1[e^{1/2(D_n-D_k)}-e^{1/2(D_n+D_k)}]F \cdot
F+z_2e^{1/2(D_k-D_n)}G \cdot H=0,\label{bf3}
\end{eqnarray}
where, as usual,  the bilinear operators $D_t$ and $\exp(\delta
D_n)$\cite{HB} are defined as:
\begin{eqnarray*}
    &&D_t^m a\cdot b \equiv\left.\left({\partial\over\partial t}-
    {\partial\over\partial t'}\right)^ma(t)b(t')\right|_{t'=t},\\
    &&\exp(\delta D_n)\;a\cdot b\equiv a(n+\delta)b(n-\delta).
\end{eqnarray*}

 In section 2, we present the
double-Casorati determinant solutions to the differential--difference system
 (\ref{N1}-\ref{N3}). Its Grammian determinant solutions  are presented in section 3, while in
section 4, a bilinear B\"{a}cklund transformation and Lax pair are
derived. Section 5 is devoted to the conclusions and a discussion
of the result obtained.

\section{Double-Casorati determinant solutions to the discrete DS system }
\setcounter{equation}{0}

It is well--known that the continuous DS equation (\ref{DS3}--\ref{DS5})
has solutions expressed in terms of  double-Wronskians of the solutions of the Spectral Problem
\cite{BLMP,F,JJ,HH,NJ}. In this section, we present the double-Casorati
determinant solutions for the  discrete DS system
(\ref{bf1}--\ref{bf3}).
An example of double-Casorati determinant solution for eqs
(\ref{N1}--\ref{N3})  is the one-soliton solution given later in Fig. 1.

Let us introduce the following double-Casorati determinant:
\beqa \nonumber
&&|0,1,\cdots,m-1;~ 0',1',\cdots,(2N-m-1)'| = \\&&\nonumber \\
&&= \left|\begin{array}{cccccc}\phi_1(n) &  \cdots & \phi_1(n+m-1);\psi_1(k) &  \cdots & \psi_1(k+2N-m-1)\\
\phi_2(n) &  \cdots & \phi_2(n+m-1);\psi_2(k) &  \cdots & \psi_2(k+2N-m-1)\\
\vdots &   &  \qquad \vdots \qquad\qquad \vdots &  &\vdots
\\
\phi_{2N}(n) &  \cdots & \phi_{2N}(n+m-1);\psi_{2N}(k) & \cdots &
\psi_{2N}(k+2N-m-1)
\end{array}
\right|, \label{CD1}
\eeqa
where $\phi_r(n,t)$ and $\psi_r(k,t)(r=1,2,\cdots,2N )$  satisfy the equations
\begin{eqnarray}
&&i\frac{\partial}{\partial t}\varphi_r(n)=-\alpha_1\varphi_r(n-1),\label{dr1}\\
&&i\frac{\partial}{\partial t}\psi_r(k)=\alpha_2\psi_r(k-1).
\label{dr2}
\end{eqnarray}
 Taking into account eq. (\ref{CD1}) we can state the following Proposition:

{\bf\large{Proposition 1 }} \textit{The  following
double-Casorati determinants
\begin{eqnarray}
&& F=|0,1,\cdots,m-1;~ 0',1',\cdots,(2N-m-1)'|,\label{cs1}\\
&& G=z_1|0,1,\cdots,m; ~0',1',\cdots,(2N-m-2)'|, \label{cs2}\\
&& H=\frac{1}{z_2}|0,1,\cdots,m-2;
~0',1',\cdots,(2N-m)'|,\label{cs3}
\end{eqnarray}
provide solutions to eqs.
(\ref{bf1}--\ref{bf3}). } \\

{\bf Proof:} From eqs. (\ref{cs1}--\ref{cs3}) for any integer number $i$ and $j$ we have
\begin{eqnarray}
&&F_{n+i,m+j}=|1,2,\cdots,m-1+i;~ 0',1',\cdots,(2N-m-1+j)'|,\label{d1}\\
&&G_{n+i,k+j}=z_1|0,1,\cdots,m+i;~ 1',2',\cdots,(2N-m-2+j)'|,\label{d6}\\
&&H_{n+i,k+j}=\frac{1}{z_2}|1,2,\cdots,m-2+i;~ 0',1',\cdots,(2N-m+j)'|.\label{d8}
\end{eqnarray}
From the equations (\ref{dr1}, \ref{dr2}) we get
\beqa
&&iF_t=-\alpha_1|-1,1,\cdots,m-1;~ 0',1',\cdots,(2N-m-1)'| \nonumber\\
&&\qquad\qquad\qquad\qquad\quad+\alpha_2|0,1,\cdots,m-1;~
(-1)',1',\cdots,(2N-m-1)'|,\label{dd1}\\
&&iG_t=z_1(-\alpha_1|-1,1,\cdots,m;~0',1',\cdots,(2N-m-2)'|\nonumber\\
&&\qquad\qquad\qquad\qquad\quad+\alpha_2|0,1,\cdots,m;
~(-1)',1',\cdots,(2N-m-2)'|),\label{dd3}\\
&&iH_t=\frac{1}{z_2}(-\alpha_1|-1,1,\cdots,m-2; ~0',1',\cdots,(2N-m)'|\nonumber\\
&&\qquad\qquad\qquad\qquad\quad+\alpha_2|0,1,\cdots,m-2; ~(-1)',1',\cdots,(2N-m)'|).\label{dd2}
\eeqa
Introducing eqs. (\ref{d1}--\ref{dd2}) into  eq.
\eqref{bf1} we get the determinant identity \cite{HB}:
\begin{eqnarray} &&|1,2,\cdots,m;~
0',1',\cdots,(2N-m-1)'||0,1,\cdots,m-1;~1',2',\cdots,(2N-m)'|\nonumber\\
&&\qquad-|1,2,\cdots,m;~
1',2',\cdots,(2N-m)'||0,1,\cdots,m-1;~0',1',\cdots,(2N-m-1)'|\nonumber\\
&&\qquad\quad+|0,1,\cdots,m;~
1',2',\cdots,(2N-m-1)'||1,2,\cdots,m-1;~
0',1',\cdots,(2N-m)'|=0.\nonumber\\
&& \label{DI}
\end{eqnarray}
Substituting  eqs. (\ref{d1}--\ref{dd3}) into eqs. (\ref{bf2}, \ref{bf3}) we
get the equations
\begin{eqnarray*}
&&\alpha_1(|-1,0,\cdots,m-2;~ 0',1',\cdots,(2N-m-1)'||1,2,\cdots,m-1;~ 0',1',\cdots,(2N-m)'|\nonumber\\
&&\qquad-|-1,1,\cdots,m-1;~
0',1',\cdots,(2N-m-1)'||0,1,\cdots,m-2; ~0',1',\cdots,(2N-m)'|\nonumber\\
&&\qquad\quad+|-1,1,\cdots,m-2;
~0',1',\cdots,(2N-m)'||0,1,\cdots,m-1;~
0',1',\cdots,(2N-m-1)'|)\nonumber\\
&&+\alpha_2(|0,1,\cdots,m-1;~ 1',2',\cdots,(2N-m)'||0,1,\cdots,m-2;~ (-1)',0',\cdots,(2N-m-1)'|\nonumber\\
&&\qquad-|0,1,\cdots,m-2; ~(-1)',1',\cdots,(2N-m)'||0,1,\cdots,m-1;~ 0',1',\cdots,(2N-m-1)'|\nonumber\\
&&\qquad\quad+|0,1,\cdots,m-1;~
(-1)',1',\cdots,(2N-m-1)'||0,1,\cdots,m-2; ~0',1',\cdots,(2N-m)'|) = 0,\\ \\
&&\alpha_1(|-1,0,\cdots,m-1;~ 0',1',\cdots,(2N-m-2)'||1,2,\cdots,m;~ 0',1',\cdots,(2N-m-1)'|\nonumber\\
&&\qquad-|-1,1,\cdots,m;
~0',1',\cdots,(2N-m-2)'||0,1,\cdots,m-1;~ 0',1',\cdots,(2N-m-1)'|\nonumber\\
&&\qquad\quad+|-1,1,\cdots,m-1;~
0',1',\cdots,(2N-m-1)'||0,1,\cdots,m; ~0',1',\cdots,(2N-m-2)'|)\nonumber\\
&&+\alpha_2(|0,1,\cdots,m;~ 1',2',\cdots,(2N-m-1)'||0,1,\cdots,m-1;~ (-1)',0',\cdots,(2N-m-2)'|\nonumber\\
&&\qquad-|0,1,\cdots,m-1;~ (-1)',1',\cdots,(2N-m-1)'||0,1,\cdots,m;
~0',1',\cdots,(2N-m-2)'|\nonumber\\
&&\qquad\quad+|0,1,\cdots,m;
~(-1)',1',\cdots,(2N-m-2)'||0,1,\cdots,m-1;~
0',1',\cdots,(2N-m-1)'|)=0,
\end{eqnarray*}
which are identically satisfied when we take into account the  determinant identities \eqref{DI}.
 In this way  Proposition 1 is
proved.

To construct  the soliton solution, we choose a simple solution of eqs.  (\ref{dr1},
\ref{dr2})
\begin{equation}
\phi_r(n,t)=\sum\limits_{l=1}\limits^{2N}a_{rl}p_l^{-n}e^{i\alpha_1p_lt},\qquad\quad
\psi_r(k,t)=\sum\limits_{l=1}\limits^{2N}b_{rl}q_l^{-k}e^{-i\alpha_2q_lt},\qquad\quad
r=1,2,\cdots,2N \label{FFF}
\end{equation}
where $p_r,q_r,a_r,b_r$ are arbitrary constants. Than the one-dromion solution of the discrete DS system is obtained by setting N=1 in eq. \eqref{CD1} and choosing $\phi_r(n),\psi_r(k)~(r=1,2)$
given by eq. \eqref{FFF} with, for example,
$\alpha_1=i,~\alpha_2=-i,~z_1=z_2=1$. In such a case
we have
\begin{eqnarray}
&&F=(a_{11}b_{21}-a_{21}b_{11})p_1^{-n}q_1^{-k}e^{-(p_1+q_1)t}+(a_{12}b_{22}-a_{22}b_{12})p_2^{-n}q_2^{-k}e^{-(p_2+q_2)t}\nonumber\\
&&\qquad\qquad+(a_{11}b_{22}-a_{21}b_{12})p_1^{-n}q_2^{-k}e^{-(p_1+q_2)t}+(a_{12}b_{21}-a_{22}b_{11})p_2^{-n}q_1^{-k}e^{-(p_2+q_1)t}\\
&&G=(a_{11}a_{22}-a_{21}a_{12})p_1^{-n}p_2^{-n}(p_2^{-1}-p_1^{-1})e^{-(p_1+p_2)t}\\
&&H=(b_{11}b_{22}-b_{21}b_{12})q_1^{-k}q_2^{-k}(q_2^{-1}-q_1^{-1})e^{-(q_1+q_2)t}.
\end{eqnarray}
In Fig. 1, we  plot the 1-dromion solution
$|v|=\frac{|G|}{F},~|w|=\frac{|H|}{F}$  in the nk--plane
with $a_{11}=a_{22}=\frac{1}{2},~a_{12}=0,~a_{21}=1,~b_{11}=\frac{3}{4},~b_{12}=\frac{1}{4},~b_{21}=-\frac{1}{4},~b_{22}=0,~
  p_1=e,~ p_2=e^{-1},~ q_1=e^{2},~q_2=e^{-2}$, at the time $t=1 $.

\begin{figure*}[htp]
\begin{center}
\begin{minipage}{0.4\textwidth}
\includegraphics[width=6cm]{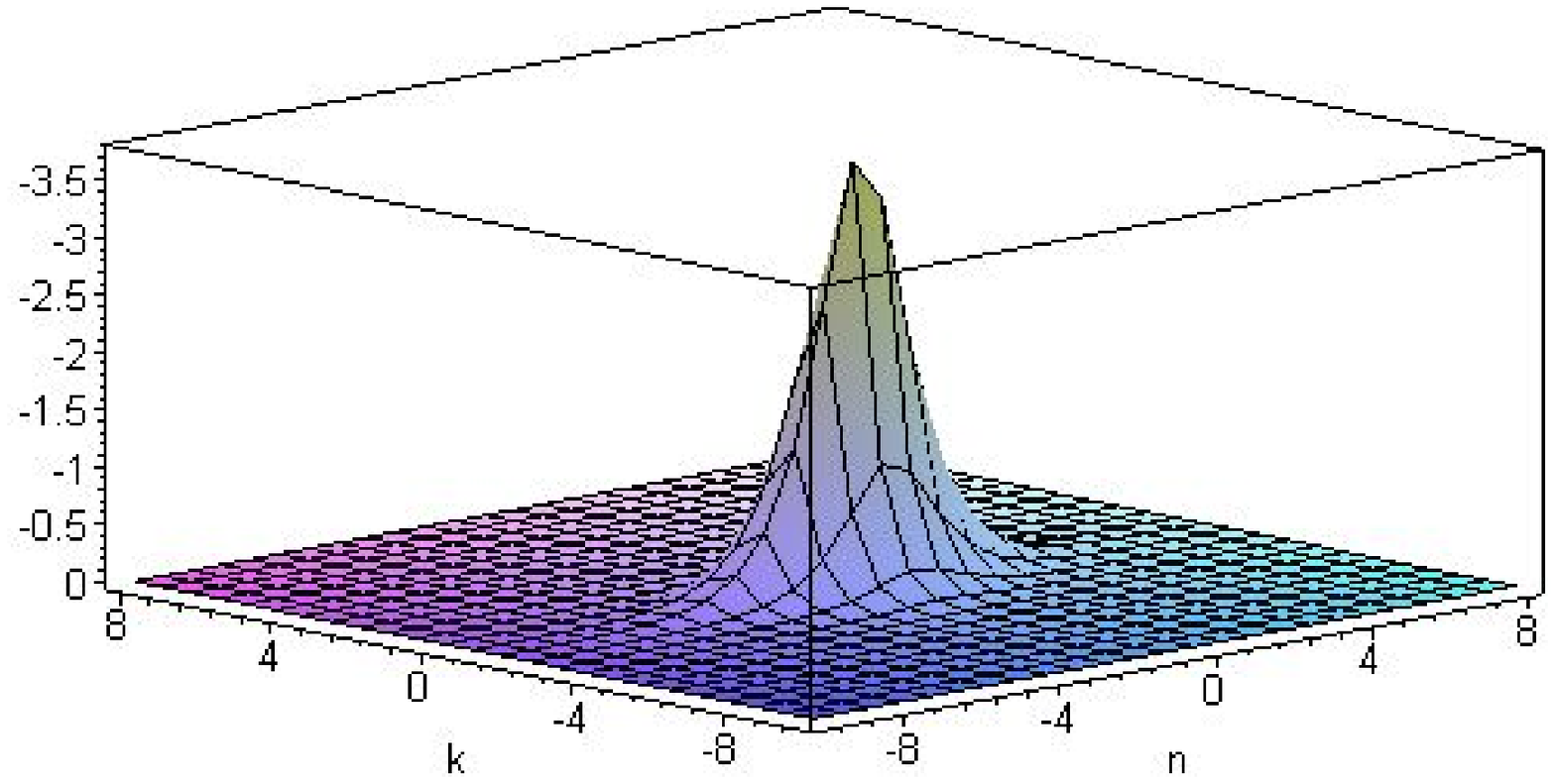}
\centerline{ $(a)$}
\end{minipage}
\hspace{.05cm}
\begin{minipage}{0.4\textwidth}
\includegraphics[width=6cm]{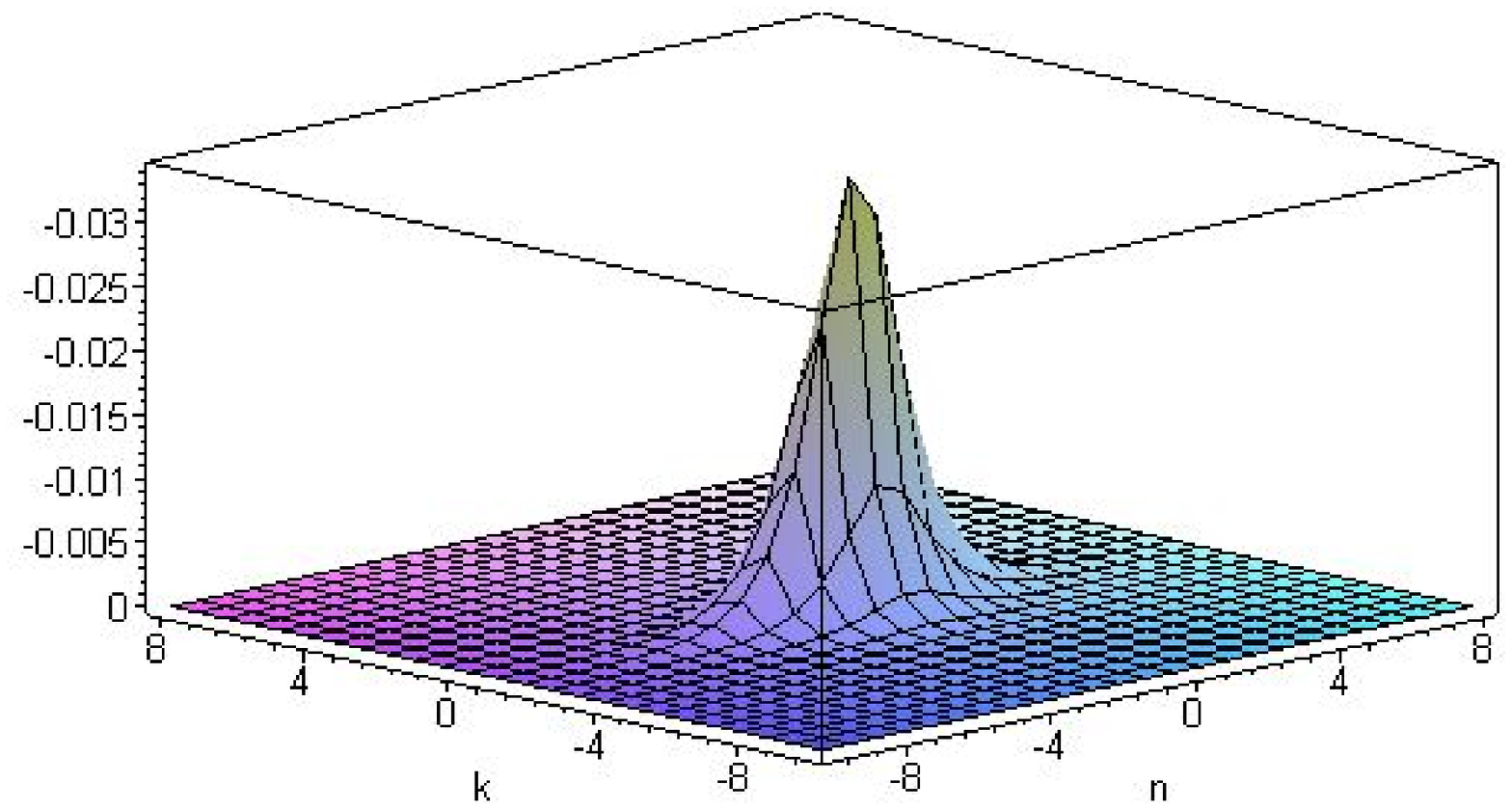}
\centerline{ $(b)$}
\end{minipage}\hspace{.05cm}
\caption{The 1-dromion solution: $(a)$ $|v|-\mbox{field }$ , $(b)$
$|w|-\mbox{field }$}
\end{center}
\end{figure*}

\section{Grammian determinant solutions to the discrete DS equation }
\setcounter{equation}{0}  The Grammian
technique was first used by Nakamura for constructing the solutions
expressed in terms of the special functions for the two-dimensional
Toda lattice equation and  the KP equation \cite{N1,N2}. In \cite{GN} we can find a Grammian
determinant solution for the continuous DS system. In this
section, we present solutions of the discrete DS system written down
in terms of Grammian determinants. At the end we show  that by a  proper
choice of parameters the double-Casorati determinant solution and
Grammian determinant solution give the same 1-soliton solution. \\

{\bf\large{Proposition 2 }}\textit{The functions  \begin{equation}
F=|C+\Omega|=|\textbf{F}|, \label{F}\end{equation}
\begin{align}\label{GH} G=\frac{z_1}{\alpha_1}\left |
\begin{array}{cc}
\textbf{F }& \Phi(n+1)\\
\Psi'(-k+1)^T  & 0
\end{array}\right|
, \qquad\qquad H=\frac{1}{z_2\alpha_2}\left|
\begin{array}{cc}
\textbf{F }& \Psi(k+1)\\
\Phi'(-n+1)^T  & 0
\end{array}\right|,
\end{align}
where $\textbf{F}$ is a
$(M+N)\times(M+N)$ matrix, $C=(c_{\mu\nu})$ is a $(M+N)\times(M+N)$ constant
matrix,  $\Omega$ is a $(M+N)\times(M+N)$ block diagonal matrix
\begin{align*} \Omega=\left (
\begin{array}{cc}
\int_{-\infty}^{t}\varphi_r(n)\varphi'_j(-n)dt & \\
 & \int_{t}^{+\infty}\psi_s(k)\psi'_l(-k)dt
\end{array}\right)
,\end{align*}
 and $\Phi, \Phi', \Psi, \Psi'$ are $M+N$ column vectors
\begin{eqnarray*}
&&\Phi(n)=(\varphi_1(n),\cdots,\varphi_{M}(n);~0,\cdots,0)^T,\\
&&\Phi'(-n)=(\varphi'_1(-n),\cdots,\varphi'_{M}(-n);~0,\cdots,0)^T,\\
&&\Psi(k)=(0,\cdots,0;~\psi_1(k),\cdots,\psi_{N}(k))^T,\\
&&\Psi'(-k)=(0,\cdots,0;~\psi_1'(-k),\cdots,\psi'_{N}(-k))^T,
\end{eqnarray*}
with $\varphi_r(n,t),\varphi'_j(n,t), \psi_s(k,t),\psi'_l(k,t)$,
$r,j\in \{1,\cdots, M\}$, $s,l\in \{1,\cdots, N\}$, satisfying the
following equations:
 \begin{eqnarray}
 &&i\frac{\partial
\varphi_r(n)}{\partial t}=-\alpha_1\varphi_r(n-1),\qquad
i\frac{\partial
\varphi'_j(-n)}{\partial t}=\alpha_1\varphi_j(-n-1),\label{gdr1}\\
&&i\frac{\partial \psi_s(k)}{\partial t}=\alpha_2\psi_s(k-1),\qquad
i\frac{\partial \psi'_l(-k)}{\partial
t}=-\alpha_2\psi'_l(-k-1),\label{gdr2}
\end{eqnarray}
solve the equations (\ref{bf1}--\ref{bf3}).}\\

{\bf Proof:} Using eqs. (\ref{gdr1}, \ref{gdr2}), we are able to
express, after some calculations,  the functions appearing in eqs.
(\ref{bf1}--\ref{bf3}) in terms of the  Grammian determinants
\begin{align}\label{eqn1} F_{n+1}=F-\frac{i}{\alpha_1}\left |
\begin{array}{cc}
\textbf{F }& \Phi(n+1)\\
\Phi'(-n)^T  & 0
\end{array}\right|,
\qquad\qquad F_{k+1}=F-\frac{i}{\alpha_2}\left |
\begin{array}{cc}
\textbf{F }& \Psi(k+1)\\
\Psi'(-k)^T  & 0 \end{array}\right|,
\end{align}

\begin{align}\label{eqn8} F_{k-1}=F+\frac{i}{\alpha_2}\left |
\begin{array}{cc}
\textbf{F }& \Psi(k)\\
\Psi'(-k+1)^T  & 0 \end{array}\right|, \quad F_{n-1}=F+\frac{i}{\alpha_1}\left |
\begin{array}{cc}
\textbf{F }& \Phi(n)\\
\Phi'(-n+1)^T  & 0 \end{array}\right|,
\end{align}

\begin{align}\label{eqn2} &F_{n+1,k+1}=F-\frac{i}{\alpha_2}\left|
\begin{array}{cc}
\textbf{F }& \Psi(k+1)\\
\Psi'(-k)^T  & 0
\end{array}\right|-\frac{i}{\alpha_1}\left|
\begin{array}{cc}
\textbf{F }& \Phi(n+1)\\
\Phi'(-n)^T  & 0
\end{array}\right|\nonumber\\
&\qquad\qquad\qquad\qquad\qquad\qquad\qquad\qquad\qquad\qquad-\frac{1}{\alpha_1\alpha_2}\left
|
\begin{array}{ccc}
\textbf{F }& \Phi(n+1) & \Psi(k+1)\\
\Phi'(-n)^T  & 0& 0\\
\Psi'(-k)^T& 0& 0
\end{array}\right|,
\end{align}

\begin{align}\label{eqn3} G_{k+1}=\frac{z_1}{\alpha_1}\left |
\begin{array}{cc}
\textbf{F }& \Phi(n+1)\\
\Psi'(-k)^T  & 0
\end{array}\right|,
\qquad\qquad H_{n+1}=\frac{1}{z_2\alpha_2}\left |
\begin{array}{cc}
\textbf{F }& \Psi(k+1)\\
\Phi'(-n)^T  & 0 \end{array}\right|,
\end{align}

\begin{align}\label{eqn7} H_{k-1}=\frac{1}{z_2\alpha_2}\left |
\begin{array}{cc}
\textbf{F }& \Psi(k)\\
\Phi'(-n+1)^T  & 0
\end{array}\right|,
\qquad\qquad G_{n-1}=\frac{z_1}{\alpha_1}\left |
\begin{array}{cc}
\textbf{F }& \Phi(n)\\
\Psi'(-k+1)^T  & 0 \end{array}\right|,
\end{align}

\begin{align}\label{eqn4} F_t=\left|
\begin{array}{cc}
\textbf{F }& \Psi(k)\\
\Psi'(-k)^T  & 0
\end{array}\right|-\left|
\begin{array}{cc}
\textbf{F }& \Phi(n)\\
\Phi'(-n)^T  & 0
\end{array}\right|,\qquad\qquad
\end{align}

\begin{align}\label{eqn6} &iG_t=\frac{z_1}{\alpha_1}\{-\alpha_2\left|
\begin{array}{cc}
\textbf{F }& \Phi(n+1)\\
\Psi'(-k)^T  & 0
\end{array}\right|-\alpha_1\left|
\begin{array}{cc}
\textbf{F }& \Phi(n)\\
\Psi'(-k+1)^T  & 0
\end{array}\right|-i\left |
\begin{array}{ccc}
\textbf{F }& \Phi(n+1) & \Phi(n)\\
\Psi'(-k+1)^T  & 0& 0\\
\Phi'(-n)^T& 0& 0
\end{array}\right|\nonumber\\
&\qquad\qquad\qquad\qquad\qquad\qquad\qquad\qquad\qquad+i\left |
\begin{array}{ccc}
\textbf{F }& \Phi(n+1) & \Psi(k)\\
\Psi'(-k+1)^T  & 0& 0\\
\Psi'(-k)^T& 0& 0
\end{array}\right|\},
\end{align}

\begin{align}\label{eqn5} &iH_t=\frac{1}{z_2\alpha_2}\{\alpha_1\left|
\begin{array}{cc}
\textbf{F }& \Psi(k+1)\\
\Phi'(-n)^T  & 0
\end{array}\right|+\alpha_2\left|
\begin{array}{cc}
\textbf{F }& \Psi(k)\\
\Phi'(-n+1)^T  & 0
\end{array}\right|+i\left |
\begin{array}{ccc}
\textbf{F }& \Psi(k+1) & \Psi(k)\\
\Phi'(-n+1)^T  & 0& 0\\
\Psi'(-k)^T& 0& 0
\end{array}\right|\nonumber\\
&\qquad\qquad\qquad\qquad\qquad\qquad\qquad\qquad\qquad-i\left |
\begin{array}{ccc}
\textbf{F }& \Psi(k+1) & \Phi(n)\\
\Phi'(-n+1)^T  & 0& 0\\
\Phi'(-n)^T& 0& 0
\end{array}\right|\}.
\end{align}

We can thus prove that the functions $F, G$ and $H$
given by eqs. (\ref{F}, \ref{GH}) effectively satisfy the discrete
DS system as,
by substituting eqs. (\ref{eqn1}--\ref{eqn8}) into eqs.
(\ref{bf1}--\ref{bf3}) we get  the following three Jacobi identities
for the determinants
\begin{align} &\left |
\begin{array}{cc}
\textbf{F }& \Phi(n+1)\\
\Phi'(-n)^T  & 0
\end{array}\right| \left|
\begin{array}{cc}
\textbf{F }& \Psi(k+1)\\
\Psi'(-k)^T  & 0
\end{array}\right| - \left|
\begin{array}{cc}
\textbf{F }
\end{array}
\right| \left |
\begin{array}{ccc}
\textbf{F }& \Phi(n+1) & \Psi(k+1)\\
\Phi'(-n)^T  & 0& 0\\
\Psi'(-k)^T& 0& 0
\end{array}\right| \nonumber\\
&\qquad\qquad\qquad\qquad -\left|
\begin{array}{cc}
 \textbf{F} & \Phi(n+1)\\
\Psi'(-k)^T & 0
\end{array}
\right| \left|
\begin{array}{ccc}
\textbf{F} & \Psi(k+1) \\
\Phi'(-n)^T & 0
\end{array}
\right|=0,\label{JI}
\end{align}
\\
\begin{align} &\{\left |
\begin{array}{cc}
\textbf{F }& \Phi(n)\\
\Phi'(-n+1)^T  & 0
\end{array}\right| \left|
\begin{array}{cc}
\textbf{F }& \Psi(k+1)\\
\Phi'(-n)^T  & 0
\end{array}\right| +\left|
\begin{array}{cc}
\textbf{F }
\end{array}
\right| \left |
\begin{array}{ccc}
\textbf{F }& \Psi(k+1) & \Phi(n)\\
\Phi'(-n+1)^T  & 0& 0\\
\Phi'(-n)^T& 0& 0
\end{array}\right| \nonumber\\
&\qquad\qquad\qquad\qquad -\left|
\begin{array}{cc}
 \textbf{F} & \Phi(n)\\
\Phi'(-n)^T & 0
\end{array}
\right| \left|
\begin{array}{ccc}
\textbf{F} & \Psi(k+1) \\
\Phi'(-n+1)^T & 0
\end{array}
\right|\}\nonumber \\
&+\{\left |
\begin{array}{cc}
\textbf{F }& \Psi(k)\\
\Psi'(-k)^T  & 0
\end{array}\right| \left|
\begin{array}{cc}
\textbf{F }& \Psi(k+1)\\
\Phi'(-n+1)^T  & 0
\end{array}\right| -\left|
\begin{array}{cc}
\textbf{F }
\end{array}
\right| \left |
\begin{array}{ccc}
\textbf{F }& \Psi(k+1) & \Psi(k)\\
\Phi'(-n+1)^T  & 0& 0\\
\Psi'(-k)^T& 0& 0
\end{array}\right| \nonumber\\
&\qquad\qquad\qquad\qquad -\left|
\begin{array}{cc}
 \textbf{F} & \Psi(k+1)\\
\Psi'(-k)^T & 0
\end{array}
\right| \left|
\begin{array}{ccc}
\textbf{F} & \Psi(k) \\
\Phi'(-n+1)^T & 0
\end{array}
\right|\}=0. \label{JII}
\end{align}
\\
\begin{align}
&\{\left |
\begin{array}{cc}
\textbf{F }& \Phi(n+1)\\
\Psi'(-k)^T  & 0
\end{array}\right| \left|
\begin{array}{cc}
\textbf{F }& \Psi(k)\\
\Psi'(-k+1)^T  & 0
\end{array}\right| +\left|
\begin{array}{cc}
\textbf{F }
\end{array}
\right| \left |
\begin{array}{ccc}
\textbf{F }& \Phi(n+1) & \Psi(k)\\
\Psi'(-k+1)^T  & 0& 0\\
\Psi'(-k)^T& 0& 0
\end{array}\right| \nonumber\\
&\qquad\qquad\qquad\qquad -\left|
\begin{array}{cc}
 \textbf{F} & \Phi(n+1)\\
\Psi'(-k+1)^T & 0
\end{array}
\right| \left|
\begin{array}{ccc}
\textbf{F} & \Psi(k) \\
\Psi'(-k)^T & 0
\end{array}
\right|\}\nonumber\\
&+\{\left |
\begin{array}{cc}
\textbf{F }& \Phi(n+1)\\
\Psi'(-k+1)^T  & 0
\end{array}\right| \left|
\begin{array}{cc}
\textbf{F }& \Phi(n)\\
\Phi'(-n)^T  & 0
\end{array}\right| -\left|
\begin{array}{cc}
\textbf{F }
\end{array}
\right| \left |
\begin{array}{ccc}
\textbf{F }& \Phi(n+1) & \Phi(n)\\
\Psi'(-k+1)^T  & 0& 0\\
\Phi'(-n)^T& 0& 0
\end{array}\right| \nonumber\\
&\qquad\qquad\qquad\qquad -\left|
\begin{array}{cc}
 \textbf{F} & \Phi(n+1)\\
\Phi'(-n)^T & 0
\end{array}
\right| \left|
\begin{array}{ccc}
\textbf{F} & \Phi(n) \\
\Psi'(-k+1)^T & 0
\end{array}
\right|\}=0. \label{JIII}
\end{align}\\

The simplest soliton solution for the discrete DS system
(\ref{N1}--\ref{N3}) is obtained by  taking the simplest possible
choice for the functions $\varphi_r,\varphi'_j,\psi_s,\psi'_l $
satisfying eqs. (\ref{gdr1}, \ref{gdr2}), i.e. an exponential
\begin{eqnarray*}
&&\varphi_r(n)=k_r^{n}e^{i\alpha_1k^{-1}_rt},\qquad \varphi'_j(-n)=\bar{k}_j^{-n}e^{-i\alpha_1\bar{k}^{-1}_jt},\\
&&\psi_s(k)=\omega_s^{k}e^{-i\alpha_2\omega^{-1}_st},\qquad
\psi'_l(-k)=\bar{\omega}_l^{-k}e^{i\alpha_2\bar{\omega}^{-1}_lt},
\end{eqnarray*}
where $k_i,\bar{k}_j,\omega_s,\bar{\omega}_l$ are arbitrary constants.\\
When $N=1$, if we take \begin{equation*} C=\left (
\begin{array}{cc}
0 & -\frac{1}{\alpha_1}\\
\frac{1}{\alpha_2} & 0
\end{array}\right),\quad k_1=\omega_1=2,\quad \bar{k}_1=(\frac{1}{2}+i)^{-1},\quad
\bar{\omega}_1=(\frac{1}{2}+i)^{-1},
\end{equation*}
 we have the following  1-soliton solution for equations \eqref{N1}-\eqref{N3}:\begin{eqnarray}
&&u=\ln(\frac{1}{\alpha_1\alpha_2}[(1+2i)^n(1+2i)^ke^{(\alpha_1-\alpha_2)t}-1]),\label{1-soliton1}\\
&&v=\alpha_2z_1\frac{2^{n+1}(\frac{1}{2}+i)^{k-1}e^{it[-\frac{1}{2}\alpha_1+(i-\frac{1}{2})\alpha_2]}}{(1+2i)^n(1+2i)^ke^{(\alpha_1-\alpha_2)t}-1},\label{1-soliton2}\\
&&w=\frac{\alpha_1}{z_2}\frac{2^{k+1}(\frac{1}{2}+i)^{n-1}e^{it[(\frac{1}{2}-i)\alpha_1+\frac{1}{2}\alpha_2]}}{(1+2i)^n(1+2i)^ke^{(\alpha_1-\alpha_2)t}-1}.\label{1-soliton3}
\end{eqnarray}
This same solution is obtained by considering the double-Casorati determinant solution (\ref{cs1}--\ref{cs3})
with $N=1$ and
\begin{eqnarray*}
&&\phi_1(n,t)=(\frac{1}{2}+i)^{-n}e^{(\frac{1}{2}+i)i\alpha_1t},\qquad
\phi_2(n,t)=(\frac{1}{2})^{-n}e^{\frac{1}{2}i\alpha_1t}\\
&&\psi_1(k,t)=(\frac{1}{2})^{-k}e^{-\frac{1}{2}i\alpha_2t},\qquad
\psi_2(k,t)=(\frac{1}{2}+i)^{-k}e^{-(\frac{1}{2}+i)i\alpha_2t}.
\end{eqnarray*}

\section{Bilinear B\"{a}cklund transformation and Lax pair}
\setcounter{equation}{0}\ \ In this section we construct a bilinear
B\"{a}cklund transformation for the bilinear equations
(\ref{bf1}--\ref{bf3}), and then we derive from it a Lax pair for the discrete
DS system (\ref{N1}--\ref{N3}).

To do so, let us redefine the functions $F$, $G$ and $H$ in term of one function $f$ depending on an additional discrete variable $m$
\begin{equation*}
F(n,k;t)=f(m,n,k;t),\qquad G(n,k;t)=f(m+1,n,k;t),\qquad
H(n,k;t)=f(m-1,n,k;t),
\end{equation*}
Then eqs. (\ref{bf1}--\ref{bf3})  can be written as:
\begin{eqnarray}
&&[iD_te^{1/2D_m}+\alpha_1e^{D_n-1/2D_m}+\alpha_2e^{D_k+1/2D_m}]f \cdot f=0,\label{nbf1}\\
&&[z_1e^{1/2(D_n-D_k)}+z_2e^{1/2(D_k-D_n)+D_m}+z_3e^{1/2(D_n+D_k)}
]f \cdot f=0.\label{nbf2}
\end{eqnarray}
We can now state the following proposition:

{\bf\large{Proposition 3}} \textit{ The bilinear system (\ref{nbf1},
\ref{nbf2}) has the B\"{a}cklund transformation
\begin{eqnarray}
&&[\beta_1e^{1/2D_n}-e^{-1/2D_n}-\mu_1e^{D_m-1/2D_n}]f \cdot
g=0,\label{BT1}\\
&&[\beta_2e^{1/2(D_m+D_k)}-e^{-1/2(D_m+D_k)}-\mu_2e^{1/2(D_m-D_k)}]f
\cdot g=0,\label{BT2}\\
 &&[iD_t-\alpha_1
\frac{\mu_1}{\beta_1}e^{D_m-D_n}-\alpha_2
\frac{\mu_2}{\beta_2}e^{-D_k}]f \cdot g=0,\label{BT3}
\end{eqnarray}where $\beta_1$, $\beta_2$, $\mu_1$, $\mu_2$ are arbitrary
constants, with  $\mu_1$, $\mu_2$ satisfying the constraint
\begin{equation}
\mu_1z_1+\mu_2z_2=0.\label{BT4}
\end{equation}} \\

{\bf Proof:} Let $f$ be a solution of equations (\ref{nbf1},
\ref{nbf2}).
Using eqs. (\ref{BT1}--\ref{BT4}), we can by straightforward
calculations show that eqs. (\ref{nbf1}, \ref{nbf2})  are satisfied for $g(m,n,k;t)$
\begin{align*}
&-[e^{1/2D_m}f\cdot f][iD_te^{1/2D_m}+\alpha_1e^{D_n-1/2D_m}+\alpha_2e^{D_k+1/2D_m}]g \cdot g \\
&\qquad\equiv
\{[iD_te^{1/2D_m}+\alpha_1e^{D_n-1/2D_m}+\alpha_2e^{D_k+1/2D_m}]f
\cdot f\}[e^{1/2D_m}g\cdot g]\nonumber\\
&\qquad\qquad-\{[iD_te^{1/2D_m}+\alpha_1e^{D_n-1/2D_m}+\alpha_2e^{D_k+1/2D_m}]g
\cdot g\}[e^{1/2D_m}f\cdot f]\\
&\qquad=2\sinh(1/2D_m)(iD_tf \cdot g)\cdot fg
+2\alpha_1\sinh(1/2(D_n-D_m))(e^{1/2D_m}f \cdot g)\cdot (e^{-1/2D_m}f \cdot g)\nonumber\\
&\qquad\qquad+2\alpha_2\sinh(1/2D_k)(e^{1/2(D_k+D_m)}f
\cdot g)\cdot (e^{-1/2(D_k+D_m)}f \cdot g\\
&\qquad=2\sinh(1/2D_m)(iD_tf \cdot g)\cdot fg
+2\alpha_1\sinh(1/2(D_n-D_m))(\frac{\mu_1}{\beta_1}e^{D_m-1/2D_n}f
\cdot g)\cdot (e^{-1/2D_m}f \cdot g)\nonumber\\
&\qquad\qquad+2\alpha_2\sinh(1/2D_k)(\frac{\mu_2}{\beta_2}e^{1/2(D_m-D_k)}f
\cdot g)\cdot (e^{-1/2(D_k+D_m)}f \cdot g\\
&\qquad=2\sinh(1/2D_m)(iD_tf \cdot g)\cdot fg -2\alpha_1\frac{\mu_1}{\beta_1}\sinh(1/2D_m))(e^{D_m-D_n}f \cdot g) \cdot fg\nonumber\\
&\qquad\qquad-2\alpha_2\frac{\mu_2}{\beta_2}\sinh(1/2D_m))(e^{-D_k}f
\cdot g)\cdot fg = 0,\\
&-[e^{1/2(D_n+D_k)}f\cdot f] [z_1e^{1/2(D_n-D_k)} + z_2e^{1/2(D_k-D_n)+D_m} + z_3e^{1/2(D_n+D_k)}] g \cdot g \\
&\qquad\equiv
\{[z_1e^{1/2(D_n-D_k)}+z_2e^{1/2(D_k-D_n)+D_m}+z_3e^{1/2(D_n+D_k)}
]f \cdot f\}[e^{1/2(D_n+D_k)}g \cdot g]\nonumber\\
&\qquad\qquad-\{[z_1e^{1/2(D_n-D_k)}+z_2e^{1/2(D_k-D_n)+D_m}+z_3e^{1/2(D_n+D_k)}
]g \cdot g \}[e^{1/2(D_n+D_k)}f \cdot f] \\
&\qquad=2z_1\sinh(-1/2D_k)(e^{1/2D_n}f \cdot g)\cdot(e^{-1/2D_n}f \cdot g)\nonumber\\
&\qquad\qquad+2z_2\sinh(1/2(D_m-D_n))(e^{1/2(D_m+D_k)}f \cdot g)\cdot(e^{-1/2(D_m+D_k)}f \cdot g)\\
&\qquad=-2z_1\mu_1\sinh(1/2D_k)(e^{1/2D_n}f \cdot g)\cdot(e^{D_m-1/2D_n}f \cdot g)\nonumber\\
&\qquad\qquad+2z_2\mu_2\sinh(1/2(D_m-D_n)(e^{1/2(D_m+D_k)}f \cdot g)\cdot(e^{1/2(D_m-D_k)}f \cdot g)\\
&\qquad=-2z_1\mu_1\sinh(1/2D_k)(e^{1/2D_n}f \cdot g)\cdot(e^{D_m-1/2D_n}f \cdot g)\nonumber\\
&\qquad\qquad-2z_2\mu_2\sinh(1/2D_k)(e^{1/2D_n}f \cdot g)\cdot(e^{D_m-1/2D_n}f \cdot g) = 0
\end{align*}
In this way, Proposition 3 is satisfied and eqs. (\ref{BT1}--\ref{BT4}) constitute a BT for
(\ref{nbf1}, \ref{nbf2}).\\

From the bilinear B\"{a}cklund transformation (\ref{BT1}-\ref{BT4}),
we can derive a Lax pair for the discrete DS system
(\ref{N1}-\ref{N3}).

Let us set
\begin{equation}
u=\ln f,\quad v=\frac{f_{m+1}}{f},\quad w=\frac{f_{m-1}}{f},\quad
\phi=\frac{g}{f}.\label{TR2}
\end{equation}
Under the dependent variable transformation \eqref{TR2}, the
bilinear BT
 (\ref{BT1}-\ref{BT3}) become  the
nonlinear equations:
\begin{eqnarray}
&& \beta_1\phi-\phi_{n+1}-\mu_1vw_{n+1}\phi_{m-1,n+1}=0,\label{LP1}\\
&&\beta_2w_{k-1}\phi_{m-1,k-1}-w_{k-1}\phi-\mu_2w\phi_{m-1}=0,\label{LP2}\\
&&
i\phi_t+\alpha_1\frac{\mu_1}{\beta_1}v_{n-1}w_{n+1}e^{u_{n+1}+u_{n-1}-2u}\phi_{m-1.n+1}+\alpha_2\frac{\mu_2}{\beta_2}e^{u_{k+1}+u_{k-1}-2u}\phi_{k+1}=0,\label{LP3}
\end{eqnarray}
where $\beta_1$, $\beta_2$, $\mu_1$, $\mu_2$ are arbitrary
constants satisfying the constraint (\ref{BT4}). Eliminating
$\phi_{m-1,n+1}$,$~\phi_{m-1,k-1}$,$~\phi_{m-1}$ from eqs.
(\ref{LP1}--\ref{LP3}), we obtain the following Lax pair
 for the differential--difference DS system (\ref{N1}--\ref{N3})
\begin{eqnarray}
&&
\beta_2(\frac{\beta_1\phi_{n-1,k-1}-\phi_{k-1}}{\mu_1v_{n-1,k-1}})-\mu_2(\frac{\beta_1\phi_{n-1}-\phi}{\mu_1v_{n-1}})-\phi w_{k-1}=0,\label{L1}\\
&&i\phi_t+\frac{\alpha_1}{\beta_1}v_{n-1}e^{u_{n-1}+u_{n+1}-2u}\frac{\beta_1\phi-\phi_{n+1}}{v}+\frac{\alpha_2}{\beta_2}\mu_2e^{u_{k-1}+u_{k+1}-2u}\phi_{k+1}=0.\label{L2}
\end{eqnarray}
By  imposing the compatibility of eqs.
(\ref{L1}, \ref{L2}) we obtain  the discrete
Davey-Stewartson system (\ref{N1}--\ref{N3}). In fact, from eq. \eqref{L1}, we can
derive
\begin{eqnarray}
&& \beta_1\phi_{n-1,k-1}=\phi_{k-1}+\frac{\mu_1}{\beta_2}v_{n-1,k-1}[\phi w_{k-1}+\frac{\mu_2}{\mu_1 v_{n-1}}(\beta_1\phi_{n-1}-\phi)],\label{eq1}\\
&&\phi_{n+1,k-1}=\beta_1\phi_{k-1}-\frac{\mu_1}{\beta_2}v_{k-1}[w_{n+1,k-1}\phi_{n+1}+\frac{\mu_2}{\mu_1
v}(\beta_1\phi-\phi_{n+1})],\label{eq2}\\
&&\beta_1\phi_{n-1,k+1}=\phi_{k+1}+\frac{\mu_1}{\mu_2}v_{n-1,k+1}[\frac{\beta_2}{\mu_1
v_{n-1}}(\beta_1\phi_{n-1}-\phi)-\phi_{k+1}w],\label{eq3}
\end{eqnarray}
the  expressions of $\phi_{n-1,k-1}, \phi_{n+1,k-1},
\phi_{n-1,k+1}$  in terms of $\phi
,\phi_{n-1},\phi_{k-1},\phi_{n+1},\phi_{k+1}$.
By differentiating eq. (\ref{L1}) with respect to $t$ and substituting into it eqs. (\ref{eq1}--\ref{eq3}) we obtain an expression in terms
 of just
$\phi,\phi_{n-1}$, $\phi_{k-1}$, $\phi_{n+1}$ and $\phi_{k+1}$.  Equating to zero the coefficients of $\phi$
, $\phi_{n-1}$, $\phi_{k-1}$, $\phi_{n+1}$, and $\phi_{k+1}$ we derive  that the coefficient of
$\phi_{n-1}$ gives eq. (\ref{N1}), the coefficient of $\phi$ gives
eq. (\ref{N2}), both the coefficients of $\phi_{n+1}$ and
$\phi_{k+1}$ give  eq. (\ref{N3}) and the coefficient of
$\phi_{k-1}$ vanishes.

\section{Conclusion.}
A discrete version of the Davey-Stewartson (DSI) system is proposed
and investigated using the bilinear method. This DSI system exhibits
N-soliton solutions expressed in terms of determinants of two different types, double--Casorati and Grammians.
Moreover, we have constructed the bilinear B\"{a}cklund
transformation and derived from it its Lax pair.

A few problems are still open. Among them the most significant is
surely to find the proper reduction which gives the
Davey-Stewartson equation from the system. Moreover, since in the continuous case we have
two physically interesting cases, the DSI and DSII equations, it would also be
interesting to find the discrete version of the DSII equation.

\section*{Acknowledgements}
G. and X.B.H. work was partially supported by the National Natural
Science Foundation of China (Grant no. 10471139), CAS President
grant, the knowledge innovation program of the Institute of
Computational Math., AMSS and Hong Kong RGC Grant No.
HKBU2016/05P. D.L.  was partially supported by  PRIN Project
``SINTESI-2004'' of the  Italian Minister for  Education and
Scientific Research and from  the Projects {\sl Sistemi dinamici
nonlineari discreti: simmetrie ed integrabilit\'a} and {\it
Simmetria e riduzione di equazioni differenziali di interesse
fisico-matematico} of GNFM--INdAM. The visit of D.L. to China was supported by a Grants for Visiting Scholars/Consultants to Institutes in Developing Countries of the ICTP.

\end{document}